\documentclass[a4paper, 10pt, conference]{ieeeconf}      % Use this line for a4
\usepackage{graphicx}

\IEEEoverridecommandlockouts                              % This command is only
\title{\LARGE \bf
Expression Recognition in the Wild Using Sequence Modeling}

\author{\parbox{16cm}{\centering
    {\large Sowmya Rasipuram, Junaid Hamid Bhat and Anutosh Maitra}\\
    {\normalsize
    Accenture Technology Labs, Bangalore, India.\\
    }}
    \thanks{This work was not supported by Accenture Technology Labs, India. }% <-this % stops a space
}

\begin{document}

%\author{Ananymous Authors \\}
\author{Sowmya Rasipuram, Junaid  Hamid Bhat and Anutosh Maitra\\ Accenture Technology Labs, Bangalore, India \\}
\pagestyle{plain}
\maketitle

%%%%%%%%%%%%%%%%%%%%%%%%%%%%%%%%%%%%%%%%%%%%%%%%%%%%%%%%%%%%%%%%%%%%%%%%%%%%%%%%
\begin{abstract}
 As we exceed upon the procedures for modelling the different aspects of behaviour, expression recognition has become a key field of research in Human Computer Interactions. Expression recognition in the wild is a very interesting problem and is challenging as it involves  detailed feature extraction and heavy computation.  This paper presents the methodologies and techniques used in our contribution to  recognize different expressions i.e., neutral, anger, disgust, fear, happiness, sadness, surprise  in ABAW competition on Aff-Wild2 database. Aff-Wild2 database consists of videos in the wild labelled for seven different expressions at frame level. We used a bi-modal approach by fusing audio and visual features and train a sequence-to-sequence model that is based on Gated Recurrent Units (GRU) and Long Short Term Memory (LSTM) network. We show experimental results on validation data.  The overall accuracy of the proposed approach is 41.5 \%, which is better than the competition baseline of 37\%. 

\end{abstract}

%%%%%%%%%%%%%%%%%%%%%%%%%%%%%%%%%%%%%%%%%%%%%%%%%%%%%%%%%%%%%%In thi%%%%%%%%%%%%%%%%%%
\section{INTRODUCTION}
Affective Behavior Analysis in the Wild (ABAW) challenge consists of three tracks for predicting valence and arousal emotions, seven discrete emotions and Action Unit (AU) recognition. In this paper, we summarize our approach and results for recognizing seven discrete emotions in the wild. The database is released as a part of the challenge and annotations are provided at frame level \cite{kollias}. 

Aff-Wild2 consists of 545 videos with 2, 786, 201 frames. Some of these videos have left and right subjects and have been annotated for both separately. This is the largest audio-visual database available with intense annotations. 
It is an extension to the Aff-Wild database \cite{affwild}, \cite{cvpr}, \cite{kollias2017}. In \cite{bmvc}, the authors use a multi-task and multi-modal framework to predict seven basic expressions and report best performance of 0.63 on this database. In \cite{kollias2018_1}, the authors use adversarial networks to predict seven basic expressions. In this paper, we present our results against \cite{kollias}.

The training, validation and testing data splits have been given by the challenge separately for all tracks. In our approach, the main steps that we followed include a) Pre-processing for face detection to extract cropped and aligned face images from all videos, b) Extraction of audio features, c) Extraction of visual features, d) Training using deep learning methods. 

\subsection{Pre-processing}
The cropped and aligned images were also given as a part of the challenge. But, the number of cropped images for every video and the number of lines in the annotation file did not match for majority of the videos. Hence, we used a different face detector to extract aligned faces. In our approach, we used OpenFace, an opensource computer vision toolbox \cite{openface}. OpenFace is very robust to detecting faces when images are non-frontal or occludedd and in low illumination conditions. It uses a CNN (Convolutional Neural Network) based face detector. OpenFace outputs aligned and cropped face images of people in videos. The size of the image obtained using OpenFace is $112*112$. The number of face images from a video obtained using OpenFace and the number of lines in the annotation file matched exactly and hence these faces were used for further processing. 

Audio is extracted using ffmpeg\footnote{https://ffmpeg.org/ffmpeg.html}, an opensource tool for audio processing. The audio files are used for further audio processing block. 

\subsection{Feature Extraction}
\subsubsection{Audio}
Audio extracted using ffmpeg is split into N overlapping segments where N corresponds to the number of lines in the annotation file for each video. The overlap is kept equal to one-half split. We extracted MFCC (Mel-Frequency Cepstral Coefficients) and melspectrogram coefficients for each audio split. MFCC's form a cepstral representation where the frequency bands are not linear but distributed according to the mel-scale. The dimension of the MFCC vector is 40. Melspectrogram (mel-frequency spectrogram) is a representation of the short-term power spectrum and is of dimension 128. The MFCC coefficients and spectrogram coefficients are fused together to form a 168 dimensional vector for every audio split.  We used a sliding window method and convert them to segments of length $15$ with an overlap of $5$ for model training. We used librosa\footnote{https://librosa.github.io/librosa/feature.html} for audio feature extraction. 

% \footnotetext{https://librosa.github.io/librosa/feature.html}

\subsubsection{Video}
It is crucial to obtain a good feature representation for every aligned face image. In recent years, many deep learning models have been used for different feature representations. In this work, we used OpenFace to extract video features \cite{openface}. 
Features such as head pose, eye gaze, land mark positions and action unit intensities (AU) are obtained from OpenFace at frame level. OpenFace generates a vector of dimension 714 with all the above features. 

\begin{figure*}[ht]
    \centering
\includegraphics[width=6in,height=3in]{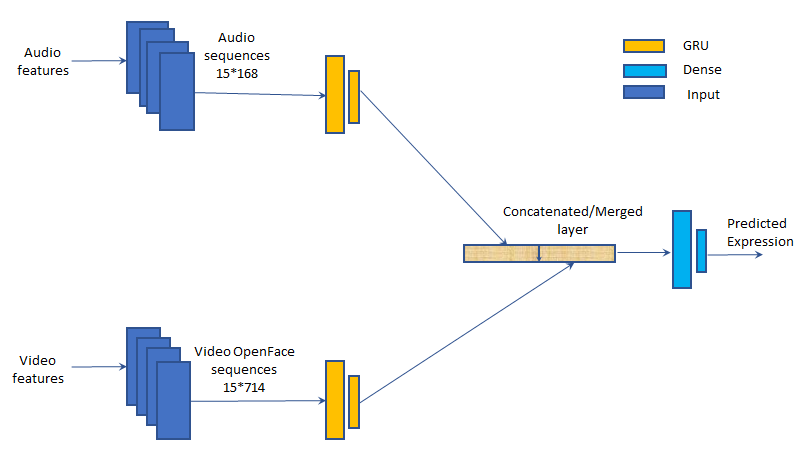}  
\caption{Block diagram of the proposed multi-modal model for Expression recognition}
    \label{fig:approach}
\end{figure*}
\subsection{Experimental Details}
Standard process of early fusion will lead to computationally expensive model and can lead to overfitting. Late and hybrid fusion methods are more suitable and prevalent. In our approach, we used a multi-layer network where we process features from each group separately and fuse them at later stages for prediction. Towards the fusion, we did several experiments within each modality and fusion. As mentioned earlier, audio features are split into overlapping segments of length 15 and the feature dimension obtained by fusing spectrogram and MFCC is 168. So, the dimension of the features from each audio sequence is 15*168.

Video features from OpenFace are also converted to segments and the dimension for every video sequence is 15*714. The two sequences are processed separately before fusion for final prediction. 

Towards fusion and testing with each modality, we performed several experiments to freeze the parameters of deep learning model. Each of the video sequences are processed separately. We implemented our codes in Python 3.7 and use Keras for deep learning framework. 
The audio sequences of dimension 15*168 are passed through two layers of GRU with 128 and 64 units respectively. We used a PReLU (Parametric ReLU) activation function with a dropout of 0.25 for each layer output. Video sequences from OpenFace of dimension 15*714 are passed through two GRU layers with dimension 256 and 64 respectively with a dropout of 0.25. All the layer outputs are followed by batch normalization.  

The 64 dimensional sequence output from the two blocks are concatenated and passed through two dense layers with 64 and 8 units respectively. The last dense layer is followed by softmax activation function. The number of units in the last dense layer is 8 as there are seven basic expressions labelled from 0-6 and video sequences with -1 from annotation file are labelled as 7. We used sparse categorical entropy with rmsporp optimizer. The learning rate has been kept at 0.0001 after performing few experiments. Figure \ref{fig:approach} represents the overall approach in a snapshot. As we use state-of-the-art methods for feature extraction that were based on deep networks and heavy training, we get promising results with our architecture. The results are validated on the validation data split provided to us. Performance metric used is the weighted sum of accuracy and F1 score as given in the white paper. 

\begin{table}[]
    \centering
       \caption{Results on validation data}
    \label{tab:results}
    \begin{tabular}{c|c|c}
    \hline
        Features & Model &  Performance\\ 
        \hline
        Baseline & & 37\% \\
        Audio only & GRU layers & 39\%  \\
        video only (OpenFace) & GRU layers & 39.9\%  \\
        Audio+Video & Figure \ref{fig:approach} & \textbf{41.5\%}  \\
        \hline
    \end{tabular} 
 
\end{table}

Table \ref{tab:results} shows experimental results for predicting performance values using audio alone, video alone and audio-visual fusion. When audio features alone are considered, we performed various experiments with networks consisting of LSTM, GRU, dense layers and our best results show a performance metric used is the weighted sum of accuracy and F1 score as given in the white paper as explained in \cite{kollias} of 37\%. When video features from expression net are considered, we observed an improvement over the baseline, performance of 39.9 \% is observed. When audio and video features are passed through the network shown in Figure \ref{fig:approach}, we observed the best performance  of 41.5\% . This indicates that fusion of features play a significant role to identify the emotion. The test results have been shared to the team for evaluation. 

\section{Conclusion}
In this paper, we proposed an architecture to automatically recognize facial expressions in the wild on Aff-Wild2 database. This framework can be used in conversational chatbot scenario where the agent can respond according to the emotion of the interactant. Our automatic framework included extraction of audio features such as MFCC's and Melspectrogram and video features that capture head pose, action unit intensities etc. An ene-to-end model is trained to perform per-frame emotion recognition using Gated Recurrent Unit. We observed significant improvement over baseline using multi-modal approach on validation data.

%%%%%%%%%%%%%%%%%%%%%%%%%%%%%%%%%%%%%%%%%%%%%%%%%%%%%%%%%%%%%%%%%%%%%%%%%%%%%%%%
\section{Acknowledgements}

The authors gratefully acknowledge the contribution of reviewers' comments.

%%%%%%%%%%%%%%%%%%%%%%%%%%%%%%%%%%%%%%%%%%%%%%%%%%%%%%%%%%%%%%%%%%%%%%%%%%%%%%%%

    \end{document}